\documentclass[aps, amsmath, twocolumn, amssymb,groupedaddress]{revtex4}
\setcitestyle{super}
\usepackage{graphicx}
\usepackage{gensymb}
\pdfoutput=1
\usepackage{amsmath}
\usepackage{color}
\usepackage{float}
\usepackage[english]{babel}
\usepackage{lipsum}
\usepackage{color,soul}

\begin{document}
\title{Discovery of highly spin-polarized conducting surface states in the strong spin-orbit coupling semiconductor Sb$_2$Se$_3$}

\author{Shekhar Das$^1$}
\author{Suman Kamboj$^1$}
\author{Anshu Sirohi$^1$}
\author{Aastha Vasdev$^1$}
\author{Sirshendu Gayen$^1$}
\author{Prasenjit Guptasarma$^2$}
\author{Goutam Sheet$^1$}
\email{goutam@iisermohali.ac.in}
\affiliation{$^1$Department of Physical Sciences,  
Indian Institute of Science Education and Research Mohali,
Sector 81, S. A. S. Nagar, Manauli, PO: 140306, India}
\affiliation{$^2$Department of Physics, University of Wisconsin, Milwaukee, Wisconsin 53211, USA}

\begin{abstract}

Majority of the A$_2$B$_3$ type chalcogenide systems with strong spin-orbit coupling, like Bi$_2$Se$_3$, Bi$_2$Te$_3$ and Sb$_2$Te$_3$ etc.,  are topological insulators. One important exception is Sb$_2$Se$_3$, where a topological non-trivial phase was argued to be possible under ambient conditions, but such a phase could be detected to exist only under pressure. In this Letter, we show that like Bi$_2$Se$_3$, Sb$_2$Se$_3$, displays generation of highly spin-polarized current under mesoscopic superconducting point contacts as measured by point contact Andreev reflection spectroscopy. In addition, we observe a large negative and anisotropic magnetoresistance in Sb$_2$Se$_3$, when the field is rotated in the basal plane. However, unlike in Bi$_2$Se$_3$, in case of Sb$_2$Se$_3$ a prominent quasiparticle interference (QPI) pattern around the defects could be obtained in STM conductance imaging. Thus, our experiments indicate that Sb$_2$Se$_3$ is a regular band insulator under ambient conditions, but due to it's high spin-orbit coupling, non-trivial spin-texture exists on the surface and the system could be on the verge of a topological insulator phase.

\end{abstract}

\maketitle


Within the band theory of solids metals and insulators are distinguished based on a band gap which is either zero for metals or non-zero for insulators. A topological insulator behaves like an insulator with a band gap in the bulk but the surface contains gapless conducting states protected by time reversal symmetry\cite{Kane, Zhang, Bernevig, Kane1}. In such systems strong spin-orbit coupling acts as an effective magnetic field pointing in a spin dependent direction thereby giving rise to non-zero spin-polarization of the conducting surface states\cite{Moore, Mele, Hsieh1}. In other words, the charge carriers corresponding to these surface states have the spin angular momentum locked with the orbital angular momentum which means carriers with definite momentum direction have definite spin. The spin polarization of the surface states of topological insulators were measured in the past by a number of techniques. The most widely exploited techniques included spin resolved ARPES \cite{Hsieh, Xu, Xia} and using circularly polarized photons to excite spin-polarized photo current \cite{McIver, Joz, Park}. Electrical methods based on fabrication of devices involving topological insulators have also been employed \cite{Li, Hong}. More recently it was shown that point contact Andreev reflection spectroscopy using a sharp tip of a conventional superconductor on the surface of a topological insulator can also be used to measure the spin polarization of the surface states through the mesurement of the degree of supression of Andreev reflection\cite{Moodera}.

Andreev reflection at an interface between a conventional superconductor and a topological material should be analyzed carefully as coupling between the superconducting order and the topological phase may be complex, particularly because of the possibility of the emergence of a topological superconductor at such interfaces\cite{Burset, Kim, Leena1, Leena2, Leena3, Shekhar}. This aspect was studied in the past and it was found that the proximity induced superconductivity in point contact geometries on topological insulators have a far greater non-topological character than topological\cite{Granstrom}. From Andreev reflection experiments on various topological insulators it was found that spin polarization in doped topological insulators can vary with the level of doping. The topological insulators Bi$_2$Te$_3$ and Sb$_2$Te$_3$ showed spin polarization of 70\% and 57\% respectively\cite{Moodera}. In all these cases, a conventional modified Blonder-Tinkham-Klapwijk (BTK) theory was used for the analysis\cite{Mazin, Strijkers}.

Based on earlier band structure calculations\cite{Vadapoo, Zhang3} Sb$_2$Se$_3$, a member of the A$_2$B$_3$ type chalcogenide family with high spin orbit coupling was categorized as a trivial band insulator under ambient conditions. Some experiments indicated the possibility of a topologically non-trivial character emerging in Sb$_2$Se$_3$ under a pressure of several giga pascals (GPa)\cite{Sood, Wang, Li1, II}. More recent band structure calculations claimed Sb$_2$Se$_3$ in fact, can be a topological insulator under ambient conditions\cite{Cao}.  In this Letter, from spin-polarized Andreev reflection spectroscopy measurements, we show that the surface of Sb$_2$Se$_3$ contains highly spin polarized surface states (with up to 70\% spin polarization) that take part in conduction leading to the generation of highly spin-polarized current. We also report detailed scanning tunneling microscopy and spectroscopy experiments showing quasiparticle interference in Sb$_2$Se$_3$ which is consistent with topologically trivial nature of Sb$_2$Se$_3$. 

\begin{figure}[h!]
	\centering
		\includegraphics[width=.485\textwidth]{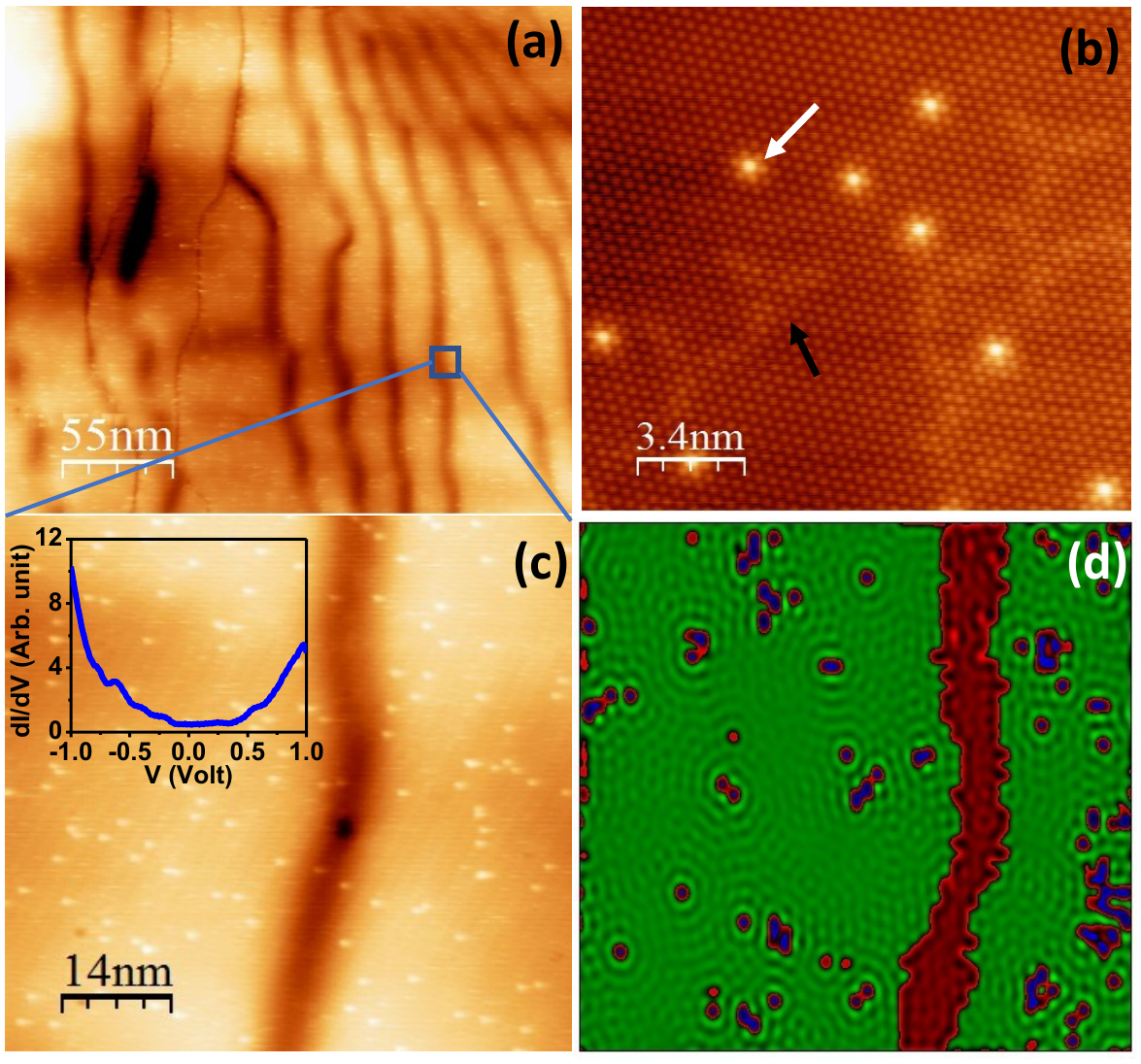}
	\caption{(a) A large area (281 nm x 281 nm) STM topograph on Sb$_2$Se$_3$ showing atomically sharp terraces. The parameters  are $ T = 17$ K, $V_s = -700$ mV and $I_s = -100$ pA (b) A representative atomic resolution image over an area of 17 nm x 17 nm on one of the terraces. (c) Image of randomly distributed defects on a selected area of 70 nm x 70 nm containing two atomically sharp steps on Sb$_2$Se$_3$. $inset$: a scanning tunneling conductance spectrum recorded away from the defects. (d) Quasi-particle interference pattern seen in the LDOS map of Sb$_2$Se$_3$ surface.}
	\label{Figure 1}
\end{figure}

\begin{figure}[h!]
	\centering
		\includegraphics[width=0.5\textwidth]{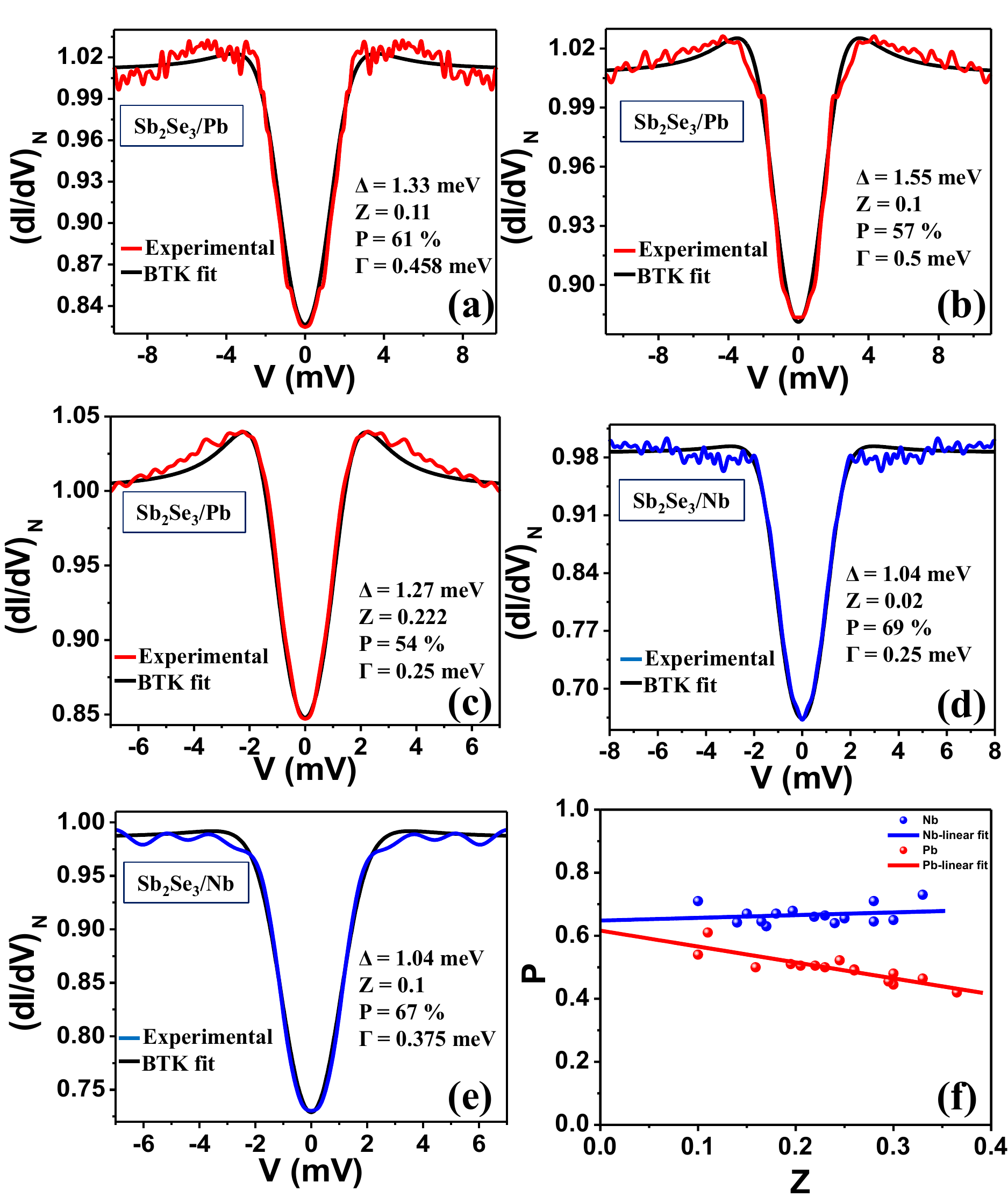}
	\caption{ Normalized $ dI/dV $ spectra for point-contacts on Sb$_2$Se$_3$ with (a,b,c) Pb tip and (d, e) Nb tip. The black lines show BTK fits with spin-polarization included. (f) Spin-polarization (P) vs. barrier strength (Z) plot. The solid lines show extrapolation to $Z = 0$ where the spin-polarization approaches 70 \%.}
	\label{Figure 2}
\end{figure}


First we confirmed the surface quality of the Sb$_2$Se$_3$ crystals using a low-temperature and ultra high vacuum scanning tunneling microscope working down to 300 mK. The crystals were cleaved $in-situ$ at 80 K under ultrahigh vacuum and were immediately transferred to the STM measurement head kept at low temperature. A large area STM image of the Sb$_2$Se$_3$ surface shows multiple extended atomic terraces with sharp steps as shown in Figure 1(a). Small area scans on top of the terraces and inside the trenches between two atomic terraces resolved atoms and the defect states (Figure 1(b)). Two types of defects are observed with one type having triangular shapes (black arrow) and the other type appear as bright spots (white arrow). The respective sides of the triangles for different triangular-shaped defects are all parallel to each other and all such defects are randomly distributed throughout the crystal surface. The triangular defects are known to be associated with Se-vacancy in the binary selenide family of materials like Bi$_2$Se$_3$\cite{Peng, Hanaguri}. The bright spots observed here might be due to Sb defects in the crystals.

In the $inset$ of Figure 1(c) we plot a typical local density of states (LDOS) spectrum recorded on Sb$_2$Se$_3$, away from the defects. The LDOS spectrum exhibits a "U"-shape with a flat bottom indicating a gap of $\sim$ 500 meV opening with the Fermi energy falling within the gap. The LDOS within the gap region does not become absolutely zero. The origin of these low-energy states is unclear at present. However, no clear signature of a surface ``Dirac cone" was observed either. We also measured the LDOS map over a selected region (Figure 1(c)) containing several defects of both kinds and a trench between two atomic terraces. As shown in Figure 1(d), such a conductance map clearly showed quasiparticle interference patterns (QPI) around the triangular and bright defects as well as across the trench. The scattering vectors responsible for QPI connect quasiparticle eigenstates of different crystal momentum \textbf{k} but with the same energy and the same spin direction. The spin selectivity, in principle, should not play a crucial role in forming the QPI pattern in a spin degenerate system. However, it is expected to significantly alter the QPI pattern on topological insulators where the spin degeneracy is lifted. In the case of Bi$_2$Se$_3$ it is seen that no prominent QPI pattern around the surface defects are seen. This absence of QPI is attributed to the mixing of the states with opposite spins in a single isotropic Dirac cone in Bi$_2$Se$_3$ through the backscattering channels\cite{Hanaguri}. Therefore, from the observation of the QPI pattern around the defects in Sb$_2$Se$_3$ it is clear that Sb$_2$Se$_3$ is a topologically trivial semiconductor. This is consistent with the initial band structure calculations and contradicts recent calculations where a topologically non-trivial phase was predicted in Sb$_2$Se$_3$ considering inter-layer van der Waals interactions. In this context, it is surprising that a large spin polarization along with anisotropic magnetic effects are observed in our field dependent Andreev reflection spectroscopy experiments as discussed below.

Point-contact Andreev reflection spectroscopy measurements\cite{BTK, Naidyuk, GoutamPRB} on Sb$_2$Se$_3$ crystals were performed using sharp tips of two conventional superconductors Pb and Nb. In Figure 2(a,b,c) and Figure 2(d,e) we show the representative Andreev reflection spectra obtained on Sb$_2$Se$_3$ with the Pb and the Nb tips respectively. The sharp dip structure at $V = 0$ along with two shallow peaks symmetric about $V = 0$ in the normalized $dI/dV$ spectra indicate considerable suppression of Andreev reflection. The black lines show the fit to the experimentally obtained spectra using Blonder-Tinkham-Klapwijk (BTK) theory modified for the finite spin-polarization of the non-superconducting electrode. The extracted values of spin-polarization $P$ are also shown. For low values of $Z$, the spin polarization is measured to be almost 70\%. For both Nb and Pb tips the extracted values of $P$ is also seen to slightly depend (linearly) on $Z$ as shown in Figure 2(f). The solid lines in Figure 2(f) show linear extrapolation of the $Z$-dependence of $P$ to $Z = 0$. This is the expected intrinsic value of the spin-polarization (for $Z = 0$). The intrinsic spin-polarization in this case is found to be approximately 65\% which is significantly large compared to some of the strong elemental ferromagnetic metals\cite{Soulen} like Fe ($ P = 40\%$), Co ($P = 42\%$) and Nickel ($P = 39\%$) and is comparable to the spin polarization of 70\% measured by Andreev reflection spectroscopy in Bi$_2$Te$_3$\cite{Moodera}. It is interesting to note that unlike in case of Bi$_2$Te$_3$, where two gap amplitudes were considered for fitting the Andreev reflection spectra, in case of Sb$_2$Se$_3$, only a single gap amplitude($\Delta$) was required which varied between 1.3 meV and 1.5 meV, as expected for Nb point contacts on a regular metal. For Pb point contacts, $\Delta$ remained comparable to the bulk gap of Pb $\sim$ 1 meV. Thus, the analysis involved only three freely varying fitting parameters --  $P$, $Z$ and $\Gamma$, the effective broadening parameter.    

The observation of high value of spin polarization in Sb$_2$Se$_3$ indicates that though the system is not categorized as a topologically nontrivial system, there are surface states with non-trivial spin-texture present due to the strong spin-orbit coupling. This supports the idea that though Sb$_2$Se$_3$ is not a topological insulator, it exists very close to a topological phase-boundary and can be driven into a topological insulator phase by small perturbation\cite{Liuu, Wang, Sood}. In the present case, the superconducting tip forming the point contact might act as a perturbation. However, it should be noted that this observation is not related to the pressure induced topological phase that was earlier observed on Sb$_2$Se$_3$ because the high spin polarization is observed even with soft tips made of superconducting Pb which cannot withstand a pressure of the order of several tens of GPa that is required to induce such a phase. Furthermore, it is known that a superconducting phase of Sb$_2$Se$_3$ is realized under high pressure\cite{Kong}, but we did not observe any signature of superconductivity of Sb$_2$Se$_3$.

\begin{figure}[h!]
	\centering
		\includegraphics[width=.5\textwidth]{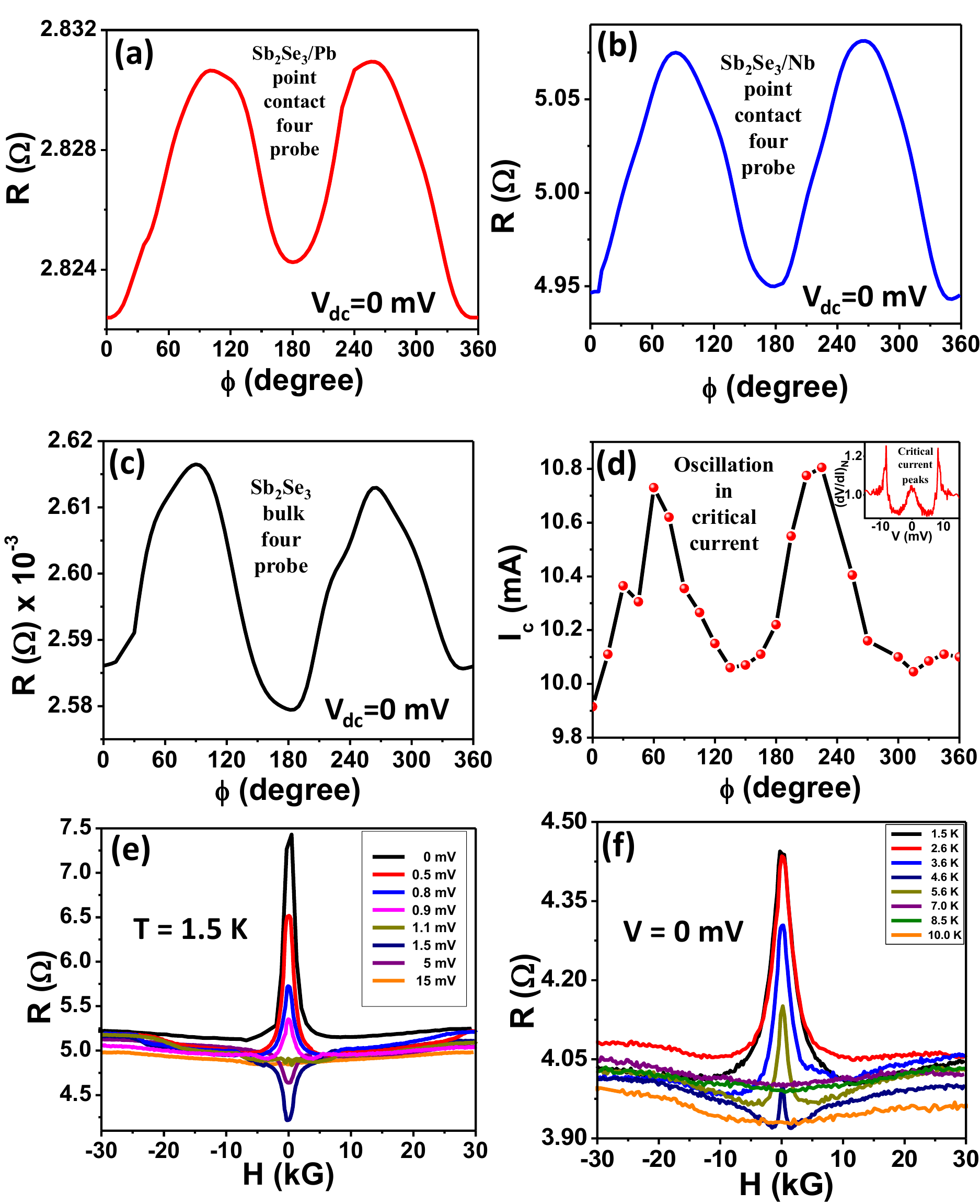}
	\caption{Anisotropic magnetoresistance (a) of a point-contact with Pb, (b) of a point-contact with Nb, (c) bulk Sb$_2$Se$_3$ crystal. (d) Anisotropy of the critical current of a superconducting point-contact in the thermal regime. The corresponding $dV/dI$ spectrum is shown in the $inset$. (e) Magneto-resistance of a Pb/Sb$_2$Se$_3$ point contact at different bias. T = 1.5 K. (f) Temperature dependence of the magnetoresistance.}
	\label{Figure 3}
\end{figure}


\begin{figure}[h!]
	\centering
		\includegraphics[width=.5\textwidth]{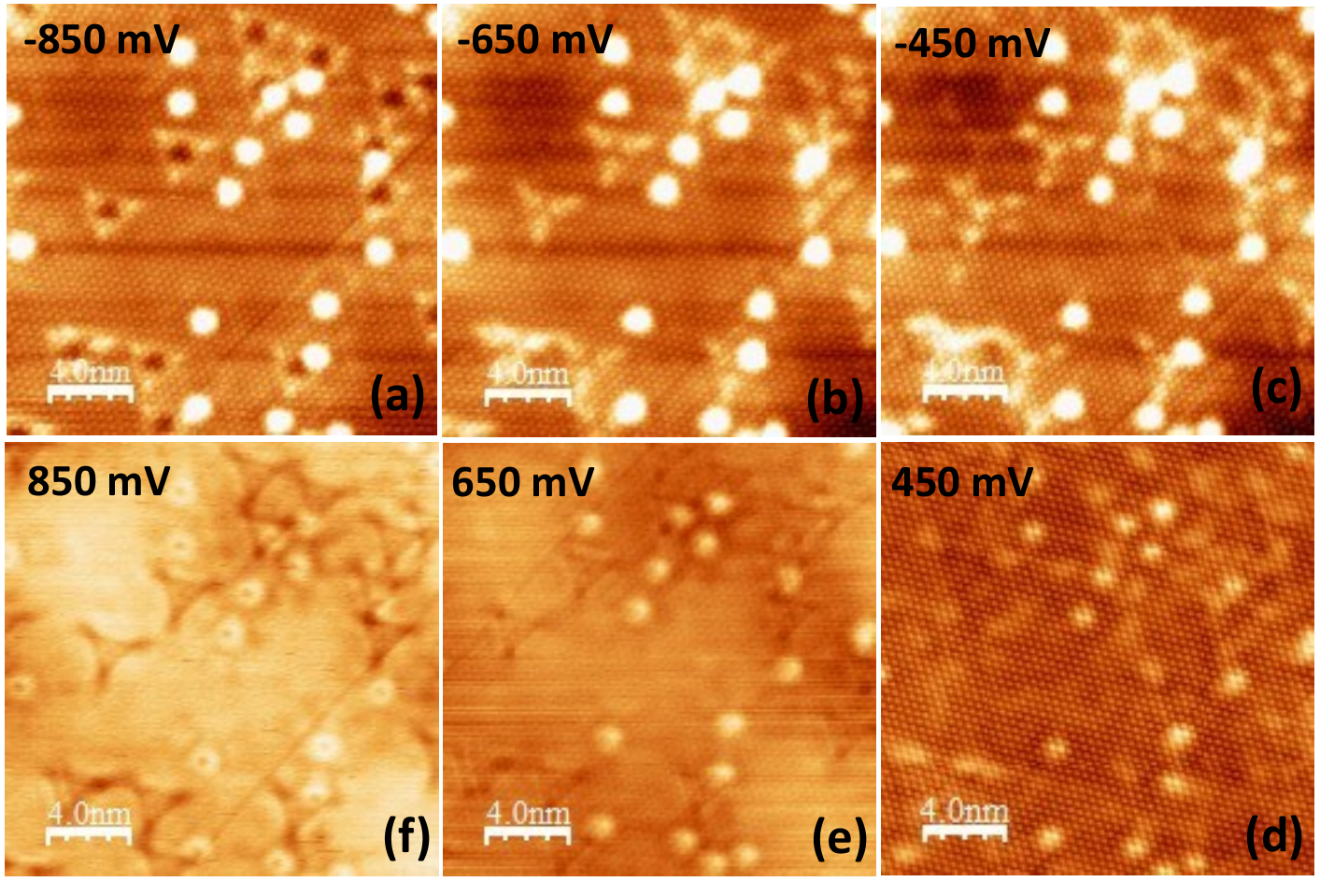}
	\caption{Evolution of STM topographs with energy. Topographs showing defect states captured at (a)-850 meV, (b) -650 meV, (c) -450 meV, (d) +450 meV, (e) +650 meV and (f) +850 meV. }
	\label{Figure 4}
\end{figure}

In order to further understand the nature of the spin-polarized surface states of Sb$_2$Se$_3$ and their implications, we measured the anisotropy of the magnetoresistance of the zero-bias point-contact resistance and the four-probe resistance of the crystal by rotating the magnetic field in the basal plane of the single crystals. We have used a 3-axis superconducting vector magnet for this measurement. As shown in Figure 3 (a) and Figure 3(b), the magnetoresistance of the point-contacts on Sb$_2$Se$_3$ with both Pb and Nb tips are seen to be highly anisotropic. We have also measured the field-angle dependence of resistance of the crystal in a conventional four-probe geometry. The four-probe resistance is also seen to be highly anisotropic confirming that the anisotropic magnetoresistance is not confined to the mesoscopic point-contacts, but originates from magnetic correlations on Sb$_2$Se$_3$ surface. The anisotropy of the magnetoresistance of the mesoscopic point-contacts and the four probe surface resistance can be attributed to the possible anisotropy of the effective spin-orbit coupling in Sb$_2$Se$_3$. Such anisotropy of the four-probe magnetoresistance was earlier observed in the topological insulator Bi$_2$Se$_3$\cite{Wang_scientific}. Therefore, the observed anisotropy also indicates that the highly spin-polarized surface states are indeed governed by the strong spin-orbit coupling in Sb$_2$Se$_3$.  

When the superconducting electrodes are in proximity of highly spin-polarized states, the critical current of the point contacts must also be modulated by the magnetic properties of such states. For investigating the modulation of critical current with magnetic field-angle, we first established point-contacts away from the ballistic regime such that Maxwell's contribution to the total point-contact resistance becomes significant and conductance dips\cite{GoutamPRB} (peaks in $dV/dI$) associated with critical current becomes prominent. Such a spectrum is shown in the $inset$ of Figure 3(d). The position of the conductance dip provides a direct measure of the critical current. After that we rotated the magnetic field in the basal plane of the crystal and recorded the spectrum for different directions of the applied magnetic field. As shown in Figure 3(d), the critical current also oscillates with the direction of the applied magnetic field in striking agreement with the anisotropic magnetoresistance presented in Figure 3(a,b,c).  

The point-contacts obtained in the ballistic regime of transport also show negative magnetoresistance. The negative magnetoresistance data is shown in Figure 3(f). The black curve shows the magnetoresistance at 1.5 K. The resistance shows a peak at zero magnetic field which is suppressed by a weak magnetic field. As the temperature is increased, the zero-field peak is systematically suppressed. This effect is completely suppressed at 7 K. In order to confirm whether the magnetoresistance is due to the superconductivity of Nb alone, the experiment was carried out at different bias applied across the tip and sample (Figure 3(e)). It is seen that with increasing the bias the peak at zero field gradually smears out and, surprisingly, becomes a dip at a bias of 1.5 mV which is the voltage corresponding to the superconducting energy gap bulk Nb. With further increasing the bias, the dip starts fading away and the magnetoresistance disappears at a bias of 15 mV. This shows that the magnetoresistance is not entirely due to the superconductivity of Nb, but Sb$_2$Se$_3$ surface also has non-trivial magnetic correlations.

It is also important to investigate any possible role of the defect states in generating the observed high spin polarization and other magnetic effects in Sb$_2$Se$_3$. In order to understand that, we carried out energy dependence of the defect states as they appear in STM topographs. In Figure 4 we show STM topographs at different bias voltages across the STM tip and the crystals in a range between -800 mV and +800 mV. It is clearly seen that the defect states change intensity and shape with energy and they are far more prominent at negative bias than at positive bias. Therefore, the defect states asymmetrically emerge and fade with respect to the direction of the applied bias. If the observed spin polarization was due to the defects, one would naturally expect the Andreev reflection spectra obtained on Sb$_2$Se$_3$ to be highly anisotropic. However, as it is seen in Figure 2, all the spectra recorded on Sb$_2$Se$_3$ are, in fact, remarkably symmetric with respect to positive and negative energy. Therefore, the role of defects behind the emergence of magnetic correlations on the surface of Sb$_2$Se$_3$ can be ruled out. 

In conclusion, we reported detailed scanning tunneling microscopy, scanning tunneling spectroscopy and point contact Andreev reflection spectroscopy on the high spin-orbit semiconductor Sb$_2$Se$_3$. We have shown the existence of two types of defects on Sb$_2$Se$_3$, one being triangular in shape as often seen in A$_2$B$_3$ type of chalcogenides. In LDOS map it was possible to obtain quasiparticle interference patterns around the defects indicating the possibility of back scattering as expected in a trivial insulator. We have further employed spin-polarized Andreev reflection spectroscopy and detected highly spin-polarized surface states in the topologically trivial band insulator Sb$_2$Se$_3$ with strong spin-orbit coupling. Furthermore, we observed highly anisotropic magnetoresistance in the basal plane of the Sb$_2$Se$_3$ crystals indicating the existence of magnetic correlations. All our observations indicate that though Sb$_2$Se$_3$ is a topologically trivial system, it possesses special surface properties unique to topological insulators and supports the idea that Sb$_2$Se$_3$ exists close to a topologically non-trivial phase due to it's high spin-orbit coupling. 

We acknowledge fruitful discussions with Umesh Waghmare. PG acknowledges support from an AFOSR-MURI grant. GS acknowledges financial support from the research grants of (a) Swarnajayanti fellowship awarded by the Department of Science and Technology (DST), Govt. of India under the grant number DST/SJF/PSA-01/2015-16, and (b) the research grant from DST-Nanomission under the grant number SR/NM/NS-1249/2013.

\end{document}